\documentclass[fleqn,usenatbib]{mnras}

\usepackage{newtxtext,newtxmath}
\usepackage[T1]{fontenc}

\DeclareRobustCommand{\VAN}[3]{#2}
\let\VANthebibliography\thebibliography
\def\thebibliography{\DeclareRobustCommand{\VAN}[3]{##3}\VANthebibliography}

\usepackage{graphicx}	
\usepackage{amsmath}	
\usepackage{cancel}

\def\beq{\begin{equation}}
\def\eeq{\end{equation}}
\def\beqa{\begin{eqnarray}}
\def\eeqa{\end{eqnarray}}
\def\bseq{\begin{subequations}\begin{eqnarray}}
\def\eseq{\end{eqnarray}\end{subequations}}

\title[Negative Power Spectrum Systematics]{The Statistics of Negative Power Spectrum Systematics in some 21~cm Analyses}

\author[Miguel F.\ Morales]{
Miguel F. Morales,$^{1}$\thanks{E-mail: miguelfm@uw.edu}
Jonathan Pober,$^{2}$
Bryna J. Hazelton$^{1,3}$
\\
$^{1}$Department of Physics, University of Washington, Seattle, USA, 98195\\
$^{2}$Department of Physics, Brown University, Providence, RI, USA, 02912\\
$^{3}$eScience Institute, University of Washington, Seattle, WA
}

\date{Accepted XXX. Received YYY; in original form ZZZ}

\pubyear{2022}

\begin{document}
\label{firstpage}
\pagerange{\pageref{firstpage}--\pageref{lastpage}}
\maketitle

\begin{abstract}
 Through a very careful analysis Kolopanis and collaborators identified a {\it negative} power spectrum (PS) systematic. The 21~cm cosmology community has assumed that any observational systematics would add power, as negative PS are non-physical. In addition to the mystery of their origin, negative PS systematics raise the spectre of artificially lowering upper limits on the 21~cm PS. It appears that the source of the negative PS systematics is a subtle interaction between choices in how the PS estimate is calculated and baseline-dependent systematic power. In this paper we present a statistical model of baseline dependent systematics to explore how negative PS systematics can appear and their statistical characteristics. This leads us to recommendations on when and how to consider negative PS systematics when reporting observational 21~cm cosmology upper limits.
\end{abstract}




\section{Introduction}

Observations of highly redshifted radio emission from neutral hydrogen have the potential to revolutionize our understanding of the Epoch of Reionization, Dark Energy, and the primordial PS at small scales \citep{ASTRO2020,Pritchard2008,Morales2010,Furlanetto2006b}. However, the signal is extraordinarily faint and separating the cosmological signal from much brighter astrophysical and anthropogenic radio foregrounds requires exquisite instrument calibration and analysis control. While all constraints on the 21~cm PS have been upper limits to date, these limits have steadily improved over the past decade. Much of this improvement has come from understanding subtle sources of `systematic'  
contamination in the measurement, calibration, and analysis. 

In a careful analysis of Murchison Widefield Array \citep{Tingay2013a} data using a delay-spectrum style analysis \citet{Kolopanis2022} identified systematics, which if not mitigated, result in a negative PS estimate. This is surprising as negative PS are non-physical---any power added to the sky must increase the observed PS---and has the potential to complicate the interpretation of upper limits. \citet{Kolopanis2022} traced the source of this negative PS to a particular type of baseline dependent systematics where antenna pairs with the same separation and orientation observe an excess systematic correlation with variable amplitude and phase. This class of systematics within nominally redundant baseline groups has become an area of particular focus in recent 21~cm cosmology measurements. 

In this paper we use a simple statistical model of baseline dependent systematics to show how this class of instrumental systematics can generate negative PS power in certain types of analysis. This directly leads to community recommendations on how to analyse future 21~cm cosmology observations and report the associated limits.

In this short paper we assume familiarity with interferometric measurements of the 21~cm PS and the associated data analysis approaches. \citet{LiuShaw2020}, \citet{Morales2010} and \citealt{Morales2019} provide overviews of modern analysis techniques and the underlying interferometric measurement process.

\section{The Source of Negative Power Spectrum Systematics}

We will start by describing how the PS is estimated in practice, with a careful focus on precisely which combinations of visibility measurements are used in the final PS estimate. 

We have individual visibility measurements of antenna pairs $v_i$. For any given separation of antennas one can have many duplicate measurements, each with different noise but observing the same underlying true signal $v_\mathbf{k}$. So our estimate of the sky signal is given by
\beq
    \hat{v}_\mathbf{k} = \langle v_i \rangle_i,
    \label{v_kestimate}
\eeq
where $\langle\ \rangle$ represents an average over the measurements $i$ (average of actual measurements, not an expectation value), and we use the subscript $\mathbf{k}$ to indicate this is of one cosmological wavenumber mode. Particularly for imaging PS this is often a weighted average, and all of the math we discuss can be generalized for the weighted case. However, much of our concern in this paper is associated with delay-spectrum type analyses of redundant arrays of near-identical antennas, where the average in Equation~\ref{v_kestimate} is effectively a straight average and can be simplified to 
\beq
    \hat{v}_\mathbf{k} = \frac{1}{N} \sum_i v_i.
\eeq

The power spectrum estimate $p_k$ then is an estimate of the variance of the observed wavenumber modes on the sky $v_\mathbf{k}$ over a selection of measurements with different $\mathbf{k}$ orientations and a narrow range of wavenumber magnitude ($k_{\rm min} < k < k_{\rm max}$)
\beq
    \hat{p}_k = \left\langle \left| \hat{v}_\mathbf{k} \right|^2 \right\rangle_\mathbf{k}.
    \label{p_est_1}
\eeq

But this is rarely used in modern analyses because the system noise on the observations is also squared creating a noise bias that must be measured and removed. Instead the estimator
\beq
    \hat{p}_k = \left\langle \hat{v}_\mathbf{k} \hat{v}_\mathbf{k}^{'*} \right\rangle_\mathbf{k},
    \label{p_est_2}
\eeq
is used where $\hat{v}_\mathbf{k}$ and $\hat{v'}_\mathbf{k}$ are matched observations of the same sky with independent noise to remove the noise bias. Matched observations are typically created by either separating the original visibilities into even \& odd time or frequency sets (a few seconds or kHz), or by re-observing the same sky and cross-multiplying measurements from different nights. 

The cross-multiplication of measurements from matched observations raises two ways of calculating $\hat{v}_\mathbf{k} \hat{v'}_\mathbf{k}^*$
\beq
    \hat{v}_\mathbf{k} \hat{v}_\mathbf{k}^{'*} = \underbrace{\frac{1}{N^2} \left( \sum_i v_i \right)\left( \sum_j v_j^{'*} \right)}_{\text{Sum-cross}} = \underbrace{\frac{1}{N^2} \sum_{ij} v_i v_j^{'*}}_{\text{Cross-sum}}.
    \label{vv_two_ways}
\eeq
In the first underbrace the individual measurements $v_i$ are summed before the conjugate multiply, while in the second underbrace all combinations of baselines are cross-multiplied then summed. While the lefthand version is the standard approach---it is computationally more efficient and maps directly to other cosmology analyses such as the CMB where noise bias is a concern---the sum-cross and cross-sum are mathematically identical and will give the same answer. 

However, the cross-sum version allows additional choices that are crucial for our story. We can sort the summation over all baseline-pairs into two sets, those that include the same antenna pair (observed at different times) and those that include only distinct antenna pairs
\beq
    \underbrace{\frac{1}{N^2} \left( \sum_i v_i \right)\left( \sum_j v_j^{'*} \right)}_{\text{Standard PS}} = \underbrace{\cancelto{}{\frac{1}{N^2} \sum_{i} v_i v_i^{'*}}}_{\text{Same-baseline term}} + \underbrace{\frac{1}{N^2} \sum_{i \ne j} v_i v_j^{'*}}_{\text{Cross-baseline term}}.
    \label{standard_vs_crossPS}
\eeq
In some implementations of the delay PS the same-baseline term is discarded and only the cross-baseline term is kept (e.g.\ \citealt{Parsons2012,Parsons2014,Kolopanis2019,HERA1Limits,Kolopanis2022}). In the absence of systematics the cross-baseline term is a perfectly fine estimator of the power spectrum. However, when baseline dependent systematics are present the standard PS form and the cross-baseline form are not equivalent, and baseline dependent systematics can source negative PS systematics in analyses that discard the same-baseline term.



\subsection{A statistical model of systematics within nominally redundant baselines}

 In the literature there is a extensive history of finding spurious `systematic' signals that appear very different on nominally redundant baselines and do not project to the sky (non-physical fringe rotation speeds). Some of these have been tracked back to over-the-air noise-coupling between antenna low noise amplifiers
 , analog and digital cross-talk \citep{Kern2019a,Kern2020a}, re-radiation of amplified sky signals \citep{HERAMemo104}, and reflected sky signals or standing waves within an antenna array \citep{Fagnoni2021,CHIME_overview}. Despite the variety of root causes, the key features we would like to model are:
\begin{itemize}
    \item The systematic is the same in the matched observations. While the systematics will differ on nominally redundant baselines ($v_i$ \& $v_j$), they appear the same in the two matched observations for a single baseline  $v_i$ \& $v'_i$. 
    \item The systematic signal does not project to the sky. Visibilities from celestial sources phase rotate based on the antenna separation and the changing location of the celestial source. The systematics seen in instruments are often very steady in time, and at a minimum have a phase rotation that is non-celestial.
    \item The systematics have apparently random phase for nominally redundant baselines. The cases described above have the fingerprint that baselines with the same antenna separation will have systematics of varying amplitude and apparently random phase, and this is often associated with changes in the far antenna sidelobes or complex reflection patterns. 
\end{itemize}
 This motivates us to statistically model our systematics as a complex random variable $s_i$ that varies per baseline (we'll revisit the per-baseline assumption in \S\ref{sec:PSpdf}). While remarkably simple---we are only assuming that there is a signal of random phase added to individual visibilities---this is sufficient to show the origin of negative PS systematics and which analyses are susceptible to this type of contamination.

For each baseline we can now add a systematic
\beq
    v_i = v_{\rm{C}} + s_i
\eeq
where $v_{\rm{C}}$ is the celestial sky signal. The key $\hat{v}_\mathbf{k} \hat{v'}_\mathbf{k}^*$ in estimating the PS can be expanded to give
\beq
    \hat{v}_\mathbf{k} \hat{v'}_\mathbf{k}^* = |v_{\rm{C}}|^2 +  \frac{1}{N^2} \left| \sum_i s_i \right|^2 + \cancelto{}{\frac{2}{N^2} \sum_i \rm{Re}(v_{\rm{C}}s_i)}.
    \label{eq:simplifying_systematics}
\eeq
Looking at the three terms on the right, the first is the PS of the celestial sky (what we fundamentally wanted), the second is the PS added by the systematic term, and the third is the sky-systematic cross-term. While the third term is interesting in its own right, we are primarily concerned with the case of $v_{\rm{C}} << s_i$, for example delay modes within the 21~cm PS window, and we can ignore the third term for this discussion. So the key concern is the statistics of the second term, and how these affect different estimates of the PS. We can then explore the PS of the systematics by substituting into Equation~\ref{standard_vs_crossPS} to obtain
\beq
    \underbrace{\frac{1}{N^2} \left| \sum_i s_i \right|^2}_{\text{Standard PS}} = \underbrace{\frac{1}{N^2} \sum_{i} |s_i|^2 }_{\text{Same-baseline term}} + \underbrace{\frac{1}{N^2} \sum_{i \ne j} s_i s_j^*}_{\text{Cross-baseline term}},
    \label{eq:systematic_statistics}
\eeq
where we have used the systematic appearing the same between matched observations to drop the $'$ and simplify ($s_i = s'_i$).
Stated simply we have the systematic contribution to observed PS for standard PS measurements (left term), the rarely used same baseline term, and the systematic contribution for delay spectra that use only cross baselines (third term).

\subsection{A worked example}
\label{sec:worked_example}
Equation~\ref{eq:systematic_statistics} only relies on the baseline dependent systematics being drawn from a distribution with random phase, but to elucidate the statistical characteristics we are going to work an example where our systematic $s_i$ is drawn from a standard complex normal distribution (unit variance complex Gaussian). This has the advantage of making some of the results analytic. We will return to discuss the more general case of non-Gaussian statistics, but none of the key results change. All we really need is that the underlying distribution of the systematics be random in phase. 

\begin{figure}
    \centering
    \includegraphics[width=\columnwidth]{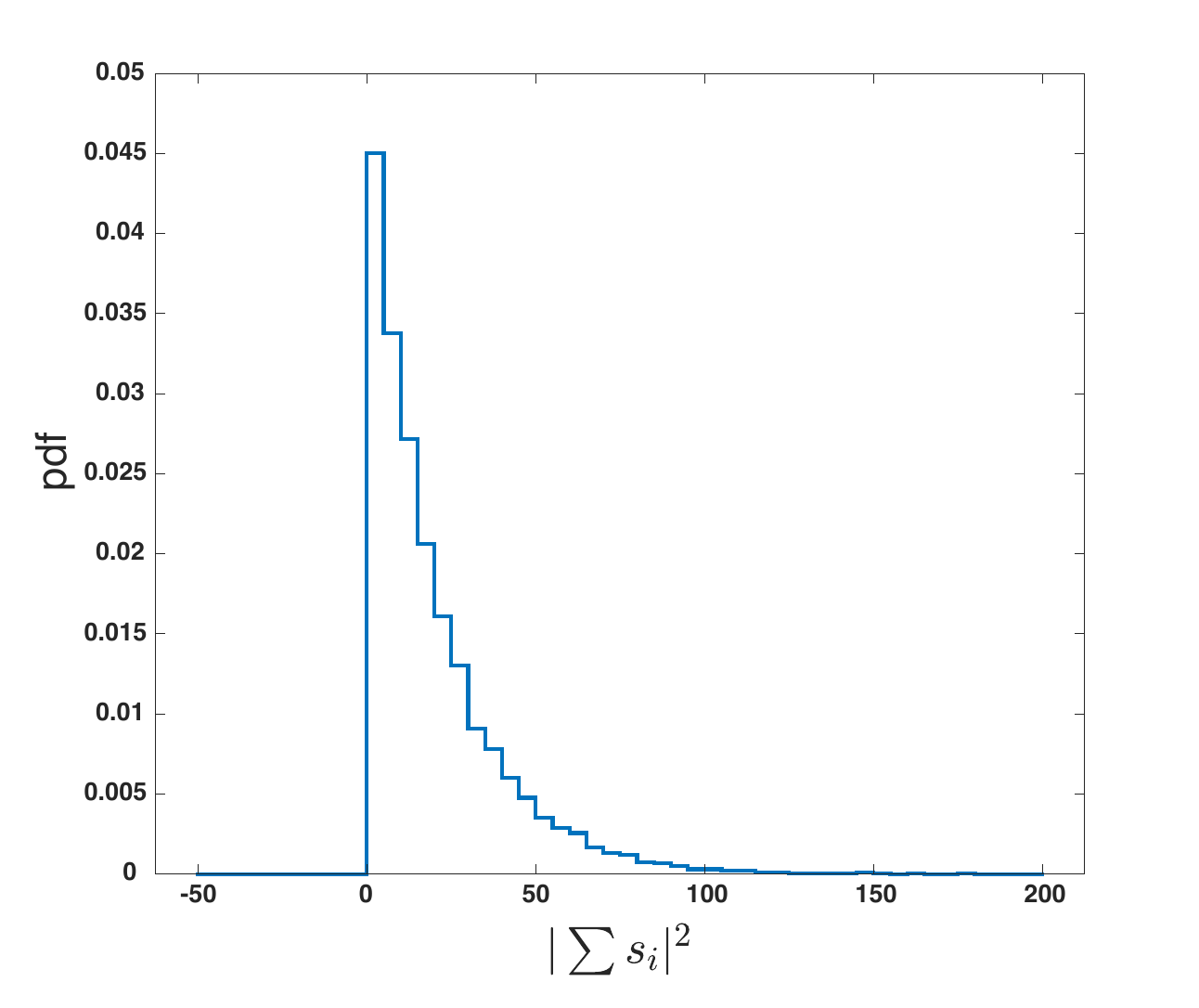}
    \caption{Probability distribution function for the standard PS term in Equation~\ref{eq:systematic_statistics} $\left| \sum_i s_i \right|^2$, for the case of 10 baselines with systematics drawn from a complex normal distribution. The resulting pdf is exponentially distributed and can only add power to a PS estimate.}
    \label{term1}
\end{figure}

\begin{figure}
    \centering
    \includegraphics[width=\columnwidth]{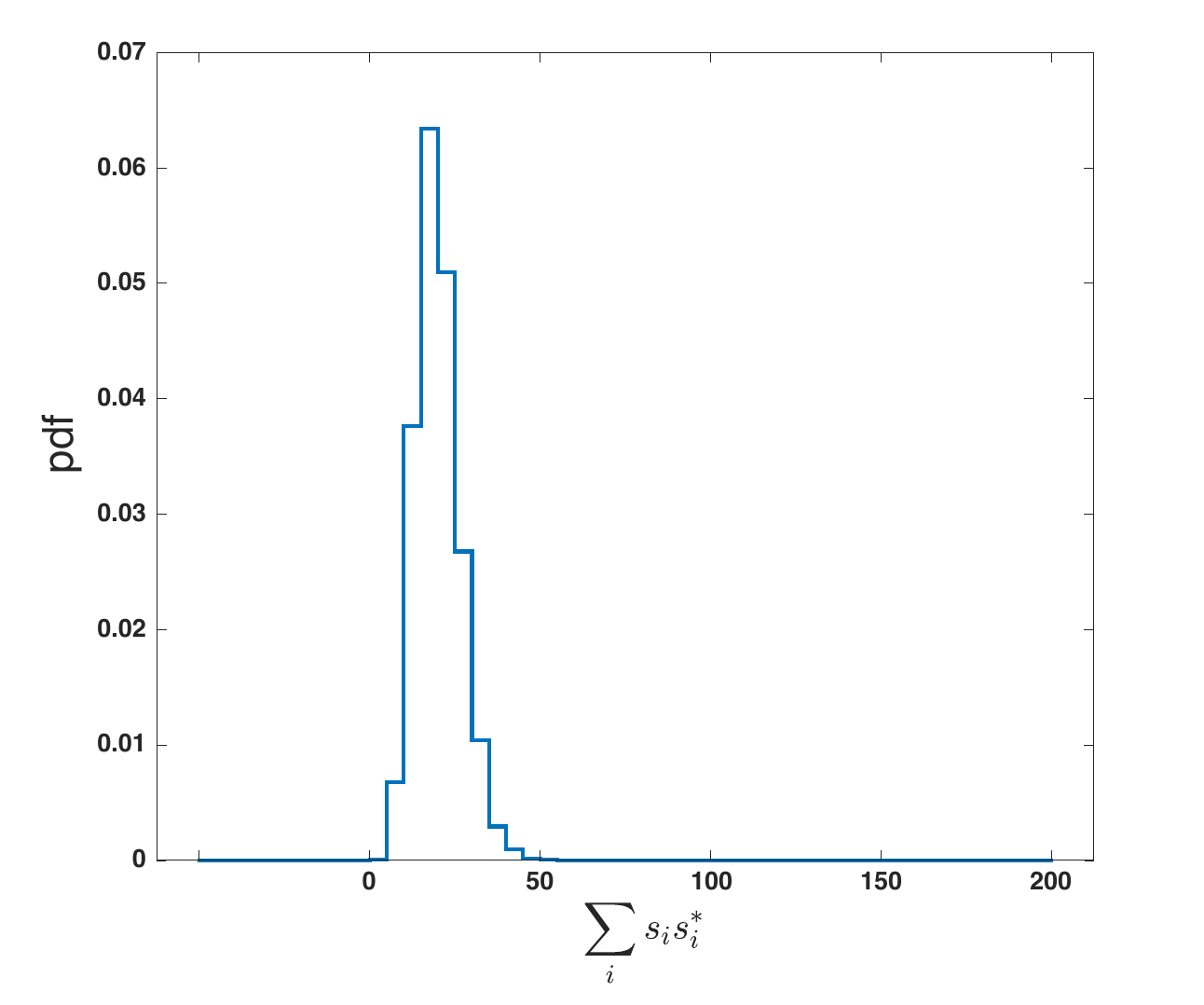}
    \caption{Probability distribution function for the same-baseline term in Equation~\ref{eq:systematic_statistics} $\sum_i |s_i|^2$, for the case of 10 baselines with systematics drawn from a complex normal distribution. The resulting pdf is the convolution of $N$ expontial distributions and is positive definite.}
    \label{term2}
\end{figure}

\begin{figure}
    \centering
    \includegraphics[width=\columnwidth]{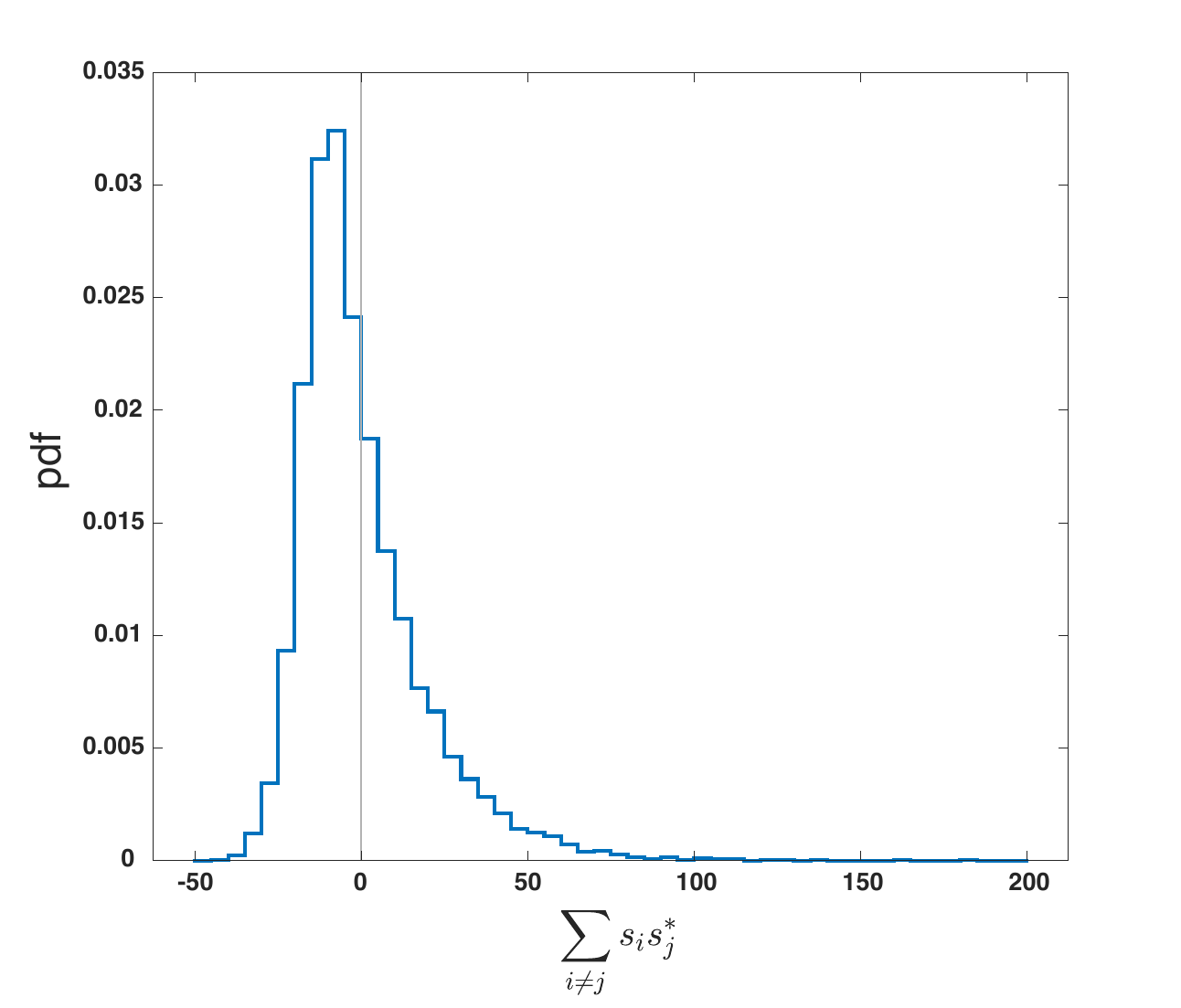}
    \caption{Probability distribution function for the cross-baseline term in Equation~\ref{eq:systematic_statistics} $\sum_{i \ne j} s_i s_j^*$, for the case of 10 baselines with systematics drawn from a complex normal distribution. The resulting pdf contains \textit{negative} PS realizations as indicated by the vertical grey line. While the systematic contribution is zero mean, $\sim$61\% of the realizations are negative and the most probable value is negative. When only the cross-term is used to estimate the PS, baseline-dependent systematics typically produce negative systematic biases in the PS.}
    \label{term3}
\end{figure}

\begin{figure}
    \centering
    \includegraphics[width=\columnwidth]{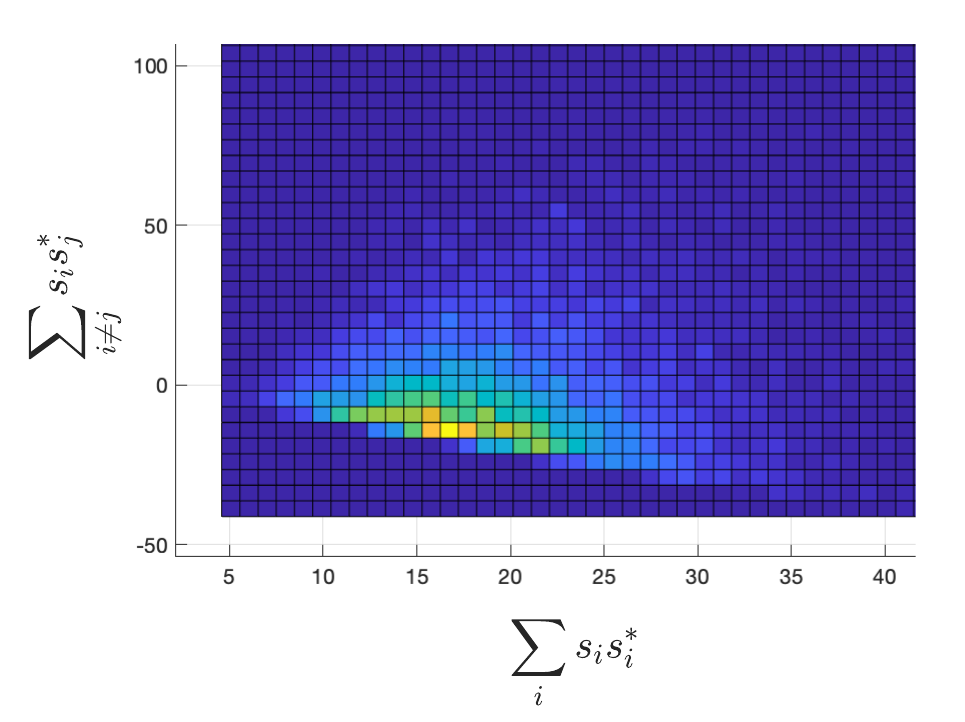}
    \caption{The joint probability of the same-baseline (horizontal) and cross-baseline (vertical) terms. Note that because the sum must be positive definite, strong negative PS realizations for the cross-baseline term necessitate positive same-baseline terms. This may provide tools for placing observational limits on the magnitude of potential negative PS. }
    \label{cross_prob}
\end{figure}

For Figures~\ref{term1}--\ref{cross_prob} we have thrown ten thousand realizations of ten baselines ($N=10$), explicitly calculated all three terms in Equation~\ref{eq:systematic_statistics} and shown the resulting probability distribution functions (pdf). For clarity we have removed the $1/N^2$ prefactor in the averages.

{\bf Standard PS $\mathbf{\left| \sum_i s_i \right|^2}$.} For the standard PS the systematics from the individual baselines are summed before squaring. The sum of complex normal variables is also complex normal, and the square of a complex normal variable is exponentially distributed \citep{BlueBook}. This can be seen in the simulated distribution in Figure~\ref{term1}. The key takeaway from this is that the systematic contribution to the standard PS is \textit{positive-definite}. The amplitude after the sum is positive.

While in theory it is possible to calculate all cross-visibilities in an imaging PS analysis, in practice it is computationally prohibitive. All of the imaging PS that have been proposed \citep{Morales2019} calculate the PS by summing first, and for them baseline-dependent systematics are purely additive. This means that the upper limits from these analyses can be interpreted in a straightforward manner.

{\bf Same-baseline term $\mathbf{\sum_i |s_i|^2}$.} While almost never used on its own, studying the pdf of the same-baseline term helps make sense of the cross-baseline term. As the square of a complex normal distribution is exponential, the analytic pdf of the second term is given by the convolution of an exponential distribution with itself $N-1$ times. This aligns with the simulated  distribution for our $N=10$ case shown in Figure~\ref{term2}. The resulting distribution is centrally peaked but positive definite and asymmetric. 

{\bf Cross-baseline term $\mathbf{\sum_{i \ne j} s_i s_j^*}$.} Figure~\ref{term3} shows the simulated pdf of the systematic PS when only cross-baseline terms are used. Interestingly the distribution is highly skewed with most of the realizations being \textit{negative}. On average the systematic PS is zero,
but the most probable PS and $\sim$61\% of the realizations are \textit{negative}. We believe this is the source of the observed negative systematics seen in delay PS analyses that only use the cross-baseline term.



Because the standard PS systematic is positive definite and equal to the sum of the same baseline and cross baseline terms for each realization, the same and cross baseline terms are not statistically independent. Figure~\ref{cross_prob} shows the joint probability distribution of the same and cross baseline terms. Strongly negative cross-baseline PS require a corresponding positive same-baseline PS so that the sum is positive definite and equal to the standard PS term.


{\bf Non-Gaussian distributions.} While the distributions for Figures~\ref{term1}--\ref{cross_prob} are specific for a complex normal distribution, the pdfs of the three terms in Equation~\ref{eq:systematic_statistics} will have the same qualitiative characteristics for non-Gaussian distributions as long as the phase of the baseline dependent systematics are random. 
The characteristic shapes of the three pdfs can be qualitatively predicted by realizing that the square of a complex random variable is necessarily positive definite. This is the fundamental reason that both the standard PS and same-baseline terms in Equation~\ref{eq:systematic_statistics} are positive definite. However, the random phase in general leads to the cross-baseline pdf to extend to negative values.  Further the cross-baseline term tends to skew negative because of the dependence between the same and cross baseline terms.


\subsection{pdf of the Power Spectrum}
\label{sec:PSpdf}
The primary quantity of interest is the distribution of systematics in the final PS estimates $\hat{p}_k$, which can be calculated from the pdfs of $\hat{v}_\mathbf{k} \hat{v'}_\mathbf{k}^*$. 
As shown in Equation~\ref{p_est_2} the PS estimate is an average over the individual $\hat{v}_\mathbf{k} \hat{v'}_\mathbf{k}^*$, so the pdf of the systematic contribution to $\hat{p}_k$ will be a repeated convolution of the $\hat{v}_\mathbf{k} \hat{v'}_\mathbf{k}^*$ distributions (Figures~\ref{term1}~or~\ref{term3}) with the number of convolutions given by the number of unique orientations and modes $\mathbf{k}$ contributing to the average. 

This has been used informally to argue that systematic contributions to the PS will be small and Gaussian. However, the observed systematics are quite large compared to the cosmological signal of interest and the number of unique systematics being averaged over is surprisingly small. Particularly for delay spectra of redundant arrays at the scientifically interesting small $|k|$ limits, the number of unique measurements can be quite small---a couple of baseline orientations and a few spectral modes. The small number of unique draws greatly reduces the averaging down of the systematic contribution to the PS, and prevents the distributions from converging to symmetric distributions. The full non-Gaussian nature of the systematic contributions has to be considered.


In addition, the number of individual systematics $s_i$ contributing to one $\hat{v}_\mathbf{k}$ may be much smaller than the number of unique baselines. In the development above we implicitly assumed that the systematic contribution to each baseline was unique, but this is an upper limit. An illustrative physical example is a systematic that is related to the characteristics of the far sidelobes of the feed, and due to manufacturing variances there are two kinds of feeds in the field. In this example the systematic contribution for each baseline would be related to the combinations of antenna feeds, with at most four unique systematic contributions. The number of draws from the random distribution is determined by the (often unknown) underlying physical cause of the systematic, and may be much smaller than the number of baselines in the sum. Even if one has many tens of baselines contributing to an average, the $i$ in Equation~\ref{eq:systematic_statistics} might be quite small. 

Other effects that are often assumed to average down random contributions don't usually apply to the systematics considered here. The systematics seen in current instruments tend to be the same for every observation of the same field. For many cases (e.g. LNA noise coupling) the systematics are steady even across fields.  While the exact pdf contribution of the systematics to the PS will depend on the instrument and observing strategy, if the systematics are modelled as being drawn from a complex random variable the number of draws is often quite small and thus non-Guassian distributions and rare outliers must be carefully considered.

\section{Discussion}

While much fainter than the astrophysical foregrounds, many recent 21~cm cosmology observations have identified baseline-dependent systematics that are significantly brighter than the cosmological signal. Of particular concern has been the identification of negative PS systematics \citep{Kolopanis2022}. In this paper we developed a simple statistical model for baseline dependent systematics, and showed how analysis choices can then lead to negative PS systematics. Because the actual causes of the systematics vary by array, treating them as a statistical ensemble is inherently limited. However, there are a number of important conclusions that we can draw.

First, all imaging PS analyses, and delay PS analyses that use all baseline pairs are immune from negative systematic contributions. This is reassuring, as physically the PS is a measure of variance and one would expect any additional systematic to add to the observed PS. This is most clearly seen in Figure~\ref{term1} where the systematic contribution is strictly positive. While it is still important for these analysis efforts to identify and remove systematics to achieve their desired sensitivity, the upper limits from these analyses can be interpreted in a straightforward manner. To form upper limits it was implicitly assumed that the systematics were additive, and in these analyses this assumption holds for the effects considered in this paper. 


In contrast, delay PS that use only the cross terms in Equation~\ref{eq:systematic_statistics} will often see negative PS contributions from baseline dependent systematics as seen in Figure~\ref{term3}. This has the unfortunate consequence of potentially \textit{subtracting} power from the cosmological PS signal. This can lead to a number of unexpected and unphysical effects, such as the detection of negative PS power in \citet{Kolopanis2022}. More concerning, it makes science interpretation of upper limits from cross baseline PS pipelines very difficult as negative systematics can artificially suppress the celestial PS signal.

So what should the community do? 

At a minimum PS measurements that use only the cross baselines need to tackle the issue of negative systematics head on through direct study of the systematics seen in that particular instrument and analysis. Unfortunately determining the properties of the systematics can be difficult---they are hard to identify and the small number of examples can make determining the underlying pdf difficult. But direct studies of the baseline dependent systematics can be used to modify reported upper limits for the potential of negative contributions to the observed PS. If the statistical distribution of the systematics is not available, since the same and cross baseline terms must add to a positive value (Figure~\ref{cross_prob}), the same baseline PS can be used to place a hard limit on how negative the systematic contribution to the observed PS can be.

Given the complexity of negative PS contributions, we would advocate that while the community is still producing PS upper limits the use of cross baseline only PS should be avoided. If all baselines are used the result is mathematically identical to the standard PS (Equation~\ref{eq:systematic_statistics}) and any systematic contribution is strictly positive. In the context of upper limits with baseline dependent systematics, proper inclusion of negative systematics removes the benefits of using only the cross baselines. The simplest response is to use all the baselines so that systematic contributions are purely additive.

As the field eventually transitions to detections and then constraints on cosmology, this context may change ($v_{\rm{C}} \gtrsim s_i$ in Equation~\ref{eq:simplifying_systematics}). The additive nature of systematics to the PS may then become a liability and the proper way of treating baseline dependent systematics may change and cross baseline analyses may experience a renaissance. But until that time the probability of baseline dependent systematics producing artificially low PS limits makes the use of cross baseline analyses problematic.

\section*{Data Availability}
No new data were generated or analysed in support of this research.

\section*{Acknowledgements}

This work was supported by NSF grants \#2107538 and \#2106510. Morales would like to thank the Murchison Widefield Array Epoch of Reionization collaboration for insightful instrumental conversations that inspired this research.



\bibliographystyle{mnras}
\bibliography{library,extra}

\begin{thebibliography}{}
\makeatletter
\relax
\def\mn@urlcharsother{\let\do\@makeother \do\$\do\&\do\#\do\^\do\_\do\%\do\~}
\def\mn@doi{\begingroup\mn@urlcharsother \@ifnextchar [ {\mn@doi@}
  {\mn@doi@[]}}
\def\mn@doi@[#1]#2{\def\@tempa{#1}\ifx\@tempa\@empty \href
  {http://dx.doi.org/#2} {doi:#2}\else \href {http://dx.doi.org/#2} {#1}\fi
  \endgroup}
\def\mn@eprint#1#2{\mn@eprint@#1:#2::\@nil}
\def\mn@eprint@arXiv#1{\href {http://arxiv.org/abs/#1} {{\tt arXiv:#1}}}
\def\mn@eprint@dblp#1{\href {http://dblp.uni-trier.de/rec/bibtex/#1.xml}
  {dblp:#1}}
\def\mn@eprint@#1:#2:#3:#4\@nil{\def\@tempa {#1}\def\@tempb {#2}\def\@tempc
  {#3}\ifx \@tempc \@empty \let \@tempc \@tempb \let \@tempb \@tempa \fi \ifx
  \@tempb \@empty \def\@tempb {arXiv}\fi \@ifundefined
  {mn@eprint@\@tempb}{\@tempb:\@tempc}{\expandafter \expandafter \csname
  mn@eprint@\@tempb\endcsname \expandafter{\@tempc}}}

\bibitem[\protect\citeauthoryear{Abdurashidova et~al.,}{Abdurashidova
  et~al.}{2022}]{HERA1Limits}
Abdurashidova Z.,  et~al., 2022, \mn@doi [The Astrophysical Journal]
  {10.3847/1538-4357/ac1c78}, 925, 221

\bibitem[\protect\citeauthoryear{Beyer}{Beyer}{1991}]{BlueBook}
Beyer W.~H.,  1991, CRC standard probability and statistics : tables and
  formulae

\bibitem[\protect\citeauthoryear{{CHIME Collaboration} et~al.,}{{CHIME
  Collaboration} et~al.}{2022}]{CHIME_overview}
{CHIME Collaboration} et~al., 2022, \mn@doi [\apjs] {10.3847/1538-4365/ac6fd9},
  \href {https://ui.adsabs.harvard.edu/abs/2022ApJS..261...29C} {261, 29}

\bibitem[\protect\citeauthoryear{Dillon, Parsons  \& Kern}{Dillon
  et~al.}{2021}]{HERAMemo104}
Dillon J.,  Parsons A.,   Kern N.,  2021, HERA Memo \#104: A Physical Model for
  the H1C Cross-Talk Systematic, \url {http://reionization.org/science/memos/}

\bibitem[\protect\citeauthoryear{Fagnoni et~al.,}{Fagnoni
  et~al.}{2021}]{Fagnoni2021}
Fagnoni N.,  et~al., 2021, \mn@doi [Monthly Notices of the Royal Astronomical
  Society] {10.1093/mnras/staa3268}, 500

\bibitem[\protect\citeauthoryear{Furlanetto, Oh  \& Briggs}{Furlanetto
  et~al.}{2006}]{Furlanetto2006b}
Furlanetto S.~R.,  Oh S.~P.,   Briggs F.~H.,  2006, \mn@doi [Physics Reports]
  {10.1016/j.physrep.2006.08.002}, 433, 181

\bibitem[\protect\citeauthoryear{Kern, Parsons, Dillon, Lanman, Fagnoni  \& de
  Lera~Acedo}{Kern et~al.}{2019}]{Kern2019a}
Kern N.~S.,  Parsons A.~R.,  Dillon J.~S.,  Lanman A.~E.,  Fagnoni N.,   de
  Lera~Acedo E.,  2019, \mn@doi [The Astrophysical Journal]
  {10.3847/1538-4357/ab3e73}, 884

\bibitem[\protect\citeauthoryear{Kern et~al.,}{Kern et~al.}{2020}]{Kern2020a}
Kern N.~S.,  et~al., 2020, \mn@doi [The Astrophysical Journal]
  {10.3847/1538-4357/ab5e8a}, 888, 70

\bibitem[\protect\citeauthoryear{Kolopanis et~al.,}{Kolopanis
  et~al.}{2019}]{Kolopanis2019}
Kolopanis M.,  et~al., 2019, \mn@doi [The Astrophysical Journal]
  {10.3847/1538-4357/ab3e3a}, 883, 133

\bibitem[\protect\citeauthoryear{Kolopanis, Pober, Jacobs  \& McGraw}{Kolopanis
  et~al.}{2022}]{Kolopanis2022}
Kolopanis M.,  Pober J.,  Jacobs D.~C.,   McGraw S.,  2022, New EoR Power
  Spectrum Limits From MWA Phase II Using the Delay Spectrum Method and Novel
  Systematic Rejection, \mn@doi{10.48550/ARXIV.2210.10885}, \url
  {https://arxiv.org/abs/2210.10885}

\bibitem[\protect\citeauthoryear{Liu \& Shaw}{Liu \& Shaw}{2020}]{LiuShaw2020}
Liu A.,  Shaw J.~R.,  2020, \mn@doi [Publications of the Astronomical Society
  of the Pacific] {10.1088/1538-3873/ab5bfd}, 132

\bibitem[\protect\citeauthoryear{Morales \& Wyithe}{Morales \&
  Wyithe}{2010}]{Morales2010}
Morales M.~F.,  Wyithe J. S.~B.,  2010, \mn@doi [Annual Review of Astronomy and
  Astrophysics] {10.1146/annurev-astro-081309-130936}, 48, 127

\bibitem[\protect\citeauthoryear{Morales, Beardsley, Pober, Barry, Hazelton,
  Jacobs  \& Sullivan}{Morales et~al.}{2019}]{Morales2019}
Morales M.~F.,  Beardsley A.,  Pober J.,  Barry N.,  Hazelton B.,  Jacobs D.,
  Sullivan I.,  2019, \mn@doi [Monthly Notices of the Royal Astronomical
  Society] {10.1093/mnras/sty2844}, 483, 2207

\bibitem[\protect\citeauthoryear{{National Academies of Sciences Engineering
  and Medicine}}{{National Academies of Sciences Engineering and
  Medicine}}{2021}]{ASTRO2020}
{National Academies of Sciences Engineering and Medicine} 2021, Pathways to
  Discovery in Astronomy and Astrophysics for the 2020s.
The National Academies Press, Washington, DC, \mn@doi{10.17226/26141}

\bibitem[\protect\citeauthoryear{Parsons, Pober, Aguirre, Carilli, Jacobs  \&
  Moore}{Parsons et~al.}{2012}]{Parsons2012}
Parsons A.~R.,  Pober J.~C.,  Aguirre J.~E.,  Carilli C.~L.,  Jacobs D.~C.,
  Moore D.~F.,  2012, \mn@doi [The Astrophysical Journal]
  {10.1088/0004-637X/756/2/165}, 756, 165

\bibitem[\protect\citeauthoryear{Parsons et~al.,}{Parsons
  et~al.}{2014}]{Parsons2014}
Parsons A.~R.,  et~al., 2014, \mn@doi [Astrophysical Journal]
  {10.1088/0004-637X/788/2/106}, 788

\bibitem[\protect\citeauthoryear{Pritchard \& Loeb}{Pritchard \&
  Loeb}{2008}]{Pritchard2008}
Pritchard J.~R.,  Loeb A.,  2008, \mn@doi [Physical Review D - Particles,
  Fields, Gravitation and Cosmology] {10.1103/PhysRevD.78.103511}, 78, 1

\bibitem[\protect\citeauthoryear{Tingay et~al.,}{Tingay
  et~al.}{2013}]{Tingay2013a}
Tingay S.~J.,  et~al., 2013, \mn@doi [Publications of the Astronomical Society
  of Australia] {10.1017/pasa.2012.007}, 30

\makeatother
\end{thebibliography}






\bsp	
\label{lastpage}
\end{document}